# The Electrochemical Transistor: a device based on the Electrochemical control of a polymer Polaronic state. The PCPDT-BT as a case study.

Andrea Stefani[1], Agnese Giacomino[2], Sara Morandi[3], Andrea Marchetti[4], Claudio Fontanesi[5,6]*

[1]*Department of Physics, FIM, Univ. of Modena, Via Campi 213/A, 41125 Modena, Italy.*
[2] *Department of Drug Science and Technology, Univ. of Torino, Via Giuria 9, 10125 Torino, Italy.*
[3] *Department of Chemistry, Univ. of Torino, Via Giuria 7, 10125 Torino, Italy.*
[4] *Department of Chemical and Geological Science, DSCG, Univ. of Modena, Via Campi 103, 41125 Modena, Italy.*
[5] *Department of Engineering "Enzo Ferrari", DIEF, Univ. of Modena, Via Vivarelli 10, 41125 Modena, Italy.*
[6] *National Interuniversity Consortium of Materials Science and Technology, INSTM, Via G. Giusti 9, 50121 Firenze (FI), Italy.*

**Abstract**

This work presents an original concept directed to implement an unconventional methodology where a device is produced by integrating a solid-state circuitry concept and an electrochemistry cell. In our experimental system an organic semiconductor, Poly[2,6-(4,4-bis-(2-ethylhexyl)-4H-cyclopenta [2,1-b;3,4-b′]dithiophene)-alt-4,7(2,1,3-benzothiadiazole)] (PCPDT-BT), serves both as the gate and working electrode. "Gating" is obtained via electrochemical polarization exploiting a conventional three electrodes electrochemical cell placed on top of a traditional source/drain/gate solid-state device configuration. Source/drain conduction is probed via impedance measurement (electrochemical impedance spectroscopy, EIS) performed as a function of time at constant frequency, under constant potential control. The conductivity of the PCPDT-BT is due to the polaronic state induced *via* application of a suitable electrochemical potential (in the oxidation regime), as it is proved by infrared (IR) spectra recorded "in-situ"/"in-operando" (in attenuated under total reflection, ATR, mode) upon both electrochemical and chemical ionization/doping of the PCPDT-BT.

Keywords: electrochemical transistor, PCPDT-BT, polaron, IRAV, infra-red, DFT

**Introduction**

The electrochemical transistor (ECT) device affirmed as a recent and intriguing paradigm. (1–3) Within this field, it is worth to mention the rising interest of "bioelectronics" oriented studies relating to the use of ECTs to be used for both disease diagnostics and therapeutics using organic electrochemical transistors (OECT).(4–6) In this field different bioelectronic devices and architectures have been developed, typically exploiting organic mixed ionic-electronic conductors (OMIEC).(7) The latter span a large number of technological applications: electrophysiological recorders(8–10), metabolite sensors(11, 12), pathogen detectors(13, 14), actuators for drug delivery, artificial muscles(15–18), also next-generation energy storage devices(19–21) just to mention a few. In these applications the device must cope with the need of intrinsic mechanical flexibility, compatibility to operate in aqueous media, and cytocompatibility.(22) Anyhow, the typical ECT scheme relates to a well-defined and coded layout, where ECTs feature a three-terminal architecture with source, drain and gate electrodes.(2, 7, 8, 23) Between the source and drain electrodes lies the channel material, through which both electronic and ionic charges are exploited as charge carriers. The gate electrodes employ in general gold or other polarizable materials, such as conjugated polymers.(24) During the operation of an ECT, a voltage bias is applied between the source and drain electrodes, driving an electronic current through the channel. The magnitude of this current is modulated by an input voltage at the gate electrode (the solution). In the traditional ECT layout the electrolyte solution serves as the "gate", being the working electrode (WE) immersed in solution and physically separated by the solid-state device layout. In this paper we introduce a substantial modification to the usual ECT experimental layout, in that in our case an organic semiconductor plays both the role of the gate and WE. Thus, the concept presented in this paper is completely different and original, the first most evident advantage relates to the organic semiconductor, whose nature and functionalization can be suitably tailored to match different applications. A distinctive feature of our geometrical layout is the possibility to carry out capacitive measurements as a function of the applied electrochemical bias, exploiting the "gate"/"WE" electrode as a capacitive sensor. PCPDT-BT organic semiconductor is used in the role of "gate"/"WE", its conductivity is determined via impedance spectroscopy measurements. Conduction in the doped state is assigned to the polaronic state, oxidation regime, as it is proved by vibrational IR spectra recorded "in-situ"/"in-operando", by using an ATR configuration, where there is evidence of the typical giant infra-red amplitude vibration (IRAV) vibrational response, i.e. the canonical vibrational polaron signature.(25–28) In addition, the de-doping kinetics, i.e. relaxation from the doped to undoped state without any applied bias, has been characterized too. Which is another original contribution to the ECT field, in that the de-doping process is not in general studied; on the contrary some interesting papers can be found concerning the doping process kinetics.(29–31) Note that, throughout this paper the term "doped polymer" refers to the presence of net positive charges injected in the polymer under electrochemical potentiostatic cotrol, i.e. PCPDT-BT oxidation process which ends up in forming "cation" PCPDT-BT units within the polymer framework.

## Results

*Electrochemical gating*

Figure 1a reports a schematic representation of our ECT experimental layout. The top part is composed by a conventional three-electrodes electrochemical cell, featuring a Pt sheet counter electrode (CE), an Ag/AgCl/KClsat reference electrode (RE) and a working electrode (WE) made of a gold pad in electrical contact with a PCPDT-BT thin film dropcasted on a solid-state device (labelled GATE in Figure1a, which is not in contact with the electrolyte solution). The solid-state device is composed of the PCPDT-BT film, which serves as the gate, drop-casted across two gold pad terminals. The latter serve as the source and drain electrical connections. The thickness of the PCPDT-BT film is in the 200 to 500 nm range, and can be controlled by changing both the concentration in the PCPDT-BT solution and the solvent. Reasonably variations in the drop casted film thickness do not influence substantially the conductivity, which is basically due to the two-dimensional (2D) PCPDT-BT film in direct contact with the solution, and in direct contact with the source/drain/gate electrodes. Figure 1b sets out the experimental configuration used for the "in-situ"/"in-operando" measurements of IR spectra, under electrochemical doping of PCPDT-BT (electrochemical oxidation regime): the doped state of a drop casted film of PCPDT-BT is controlled by applying a potential bias (electrochemical) via a glassy carbon (GC) rod kept in electrical contact with the PCPDT-BT (WE). The swallow trough immediately above the ATR crystal (zinc selenide), SPECAC HATR accessory, is exploited as the electrochemical cell: a Pt wire is the counter electrode (CE) and a Ag/AgCl wire is the reference electrode, a 0.1 M TBATFB solution in ACN is the base electrolyte.

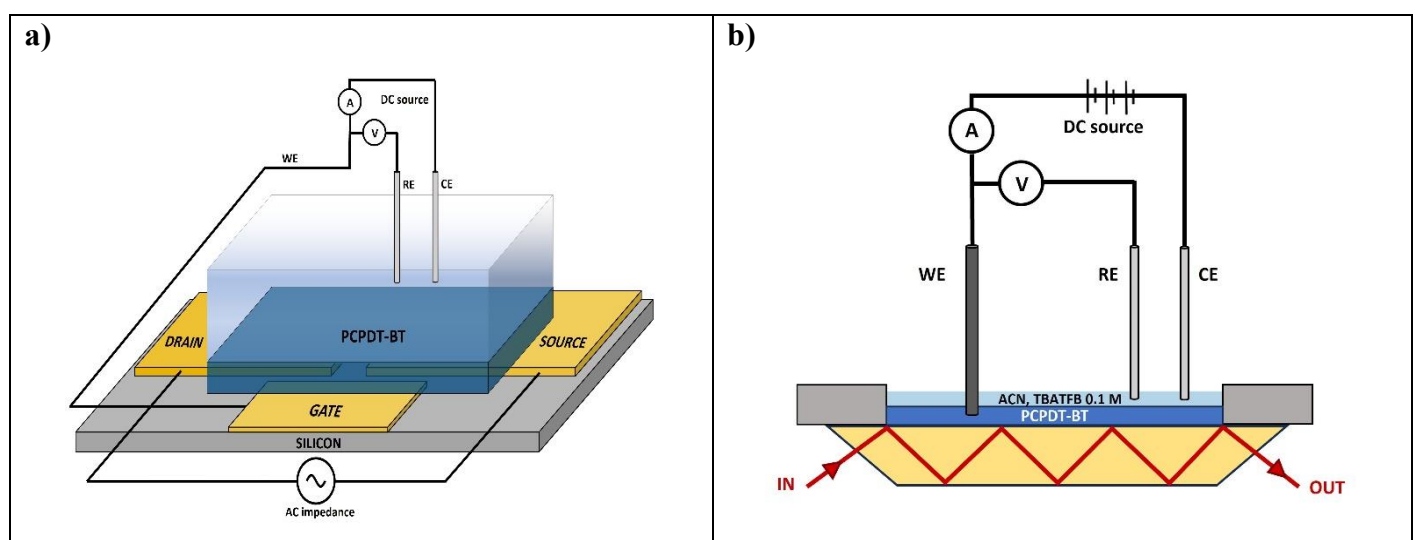

Figure 1. a) schematic representation of the electrochemical-transistor setup b) schematic representation of the experimental set up for "in-situ"/"in-operando" IR spectroelectrochemistry measurements. A thick film of PCPDT-BT acts as the WE. Ag/AgCl wire and a Pt sheet are the RE and CE, respectively. 0.1 M TBATFB is the base electrolyte in ACN.

Figure 2 shows both direct current (d.c., cyclic voltammetry, CV) and alternate current (a.c., constant frequency impedance) measurements as a function of time for the electrochemical transistor device represented in Figure 1a. Figure 2a shows the CV of a PCPDT-BT film drop casted in our ECT device: the CV is characterized by the presence of two current peaks at about +1.0 (oxidation regime) and -1.75 V (reduction regime). The relevant redox processes are of quasi-reversible nature, in that each current peak is accompanied by a "backward" peak, and the area of the "forward" and "backward" peaks are about the same (i.e., the same charge is exchanged by the forward and backward redox process). Figure 2b shows the potential applied at the gate/WE of our electrochemical transistor as a function of time, it is reported as an example the +0.1 to +0.8 V chronoamp (potentiostatic control) time pattern, with a 120 s (60 s Eini, 60 s Eox) total period. Suitable Eini and Eox potential values are selected based on the CV pattern reported in Figure 2a, to control the PCPDT-BT doped and un-doped state, oxidized and neutral respectively. A Eini = 0.1 V value was selected as the initial potential because the current is zero. Figure 2 c and d set out the resistance absolute value (Z) and the phase angle ($\phi$) as a function of time at a 97.7 Hz constant frequency, with a 5 mV peak-to-peak potential amplitude, while the PCPDT-BT un-doped/doped electronic state is controlled in real time by the electrochemical cell potential. 60 s constant potential steps are applied: gating voltage. The Eox value is varied between 0.3 to 0.9 V. Remarkably, the source to drain Z value is clearly modulated as a function of the potential applied via the electrochemical cell. A clear-cut Z variation in the resistance is evident for Eox values in between 0.6 and 0.8 V, resistance values at 0.8 and 0.9 V are almost coincident indicating a saturated doped electronic state induced in the PCPDT-BT, please compare the Figure 2c pattern. Figure 2d reports the phase vs time pattern. Thus, the overall impedance vs time outcome is in clear agreement with the CV current potential pattern. Indeed, the difference in resistance between the undoped and doped state shows a slight decrease as the frequency of the perturbation signal is increased. This result is consistent with the idea that impedance values in the low-frequency limit (d.c.) probe interfacial electrical properties. While higher frequencies probe bulk properties typical frequency range falls in between 1 to 4 kHz. In general, the frequency dependence is small in this range, while for values typically larger than 10 kHz the Debye–Falkenhagen effect takes over. Indeed, ionization in PCPDT-BT is reasonably more effective at the interface (in direct contact with the source and drain electrodes), rather than in the bulk. Remarkably, the phase values appear more sensitive to the doping of the PCPDT-BT. Figure 2b shows $\phi$ vs time transients as a function of the electrochemical potential, it seems evident a much more pronounced ability to discriminate between doped and undoped states. Focusing on the 0.7 V potential impedance vs. time series definitively the phase value at 0.7 V is closer to that of the undoped state, please compare Fig. 2b: the black line is quite close to -90°, i.e. the system behaves almost as an ideal capacitor. Then, for less positive Eox values, $\phi$ shows an almost constant value, $45 < \phi < 65$ degrees, for 0.5, 0.6, 0.7 V respectively, revealing a clear-cut discontinuity in $\phi$ between 0.7 and 0.8 V degrees.

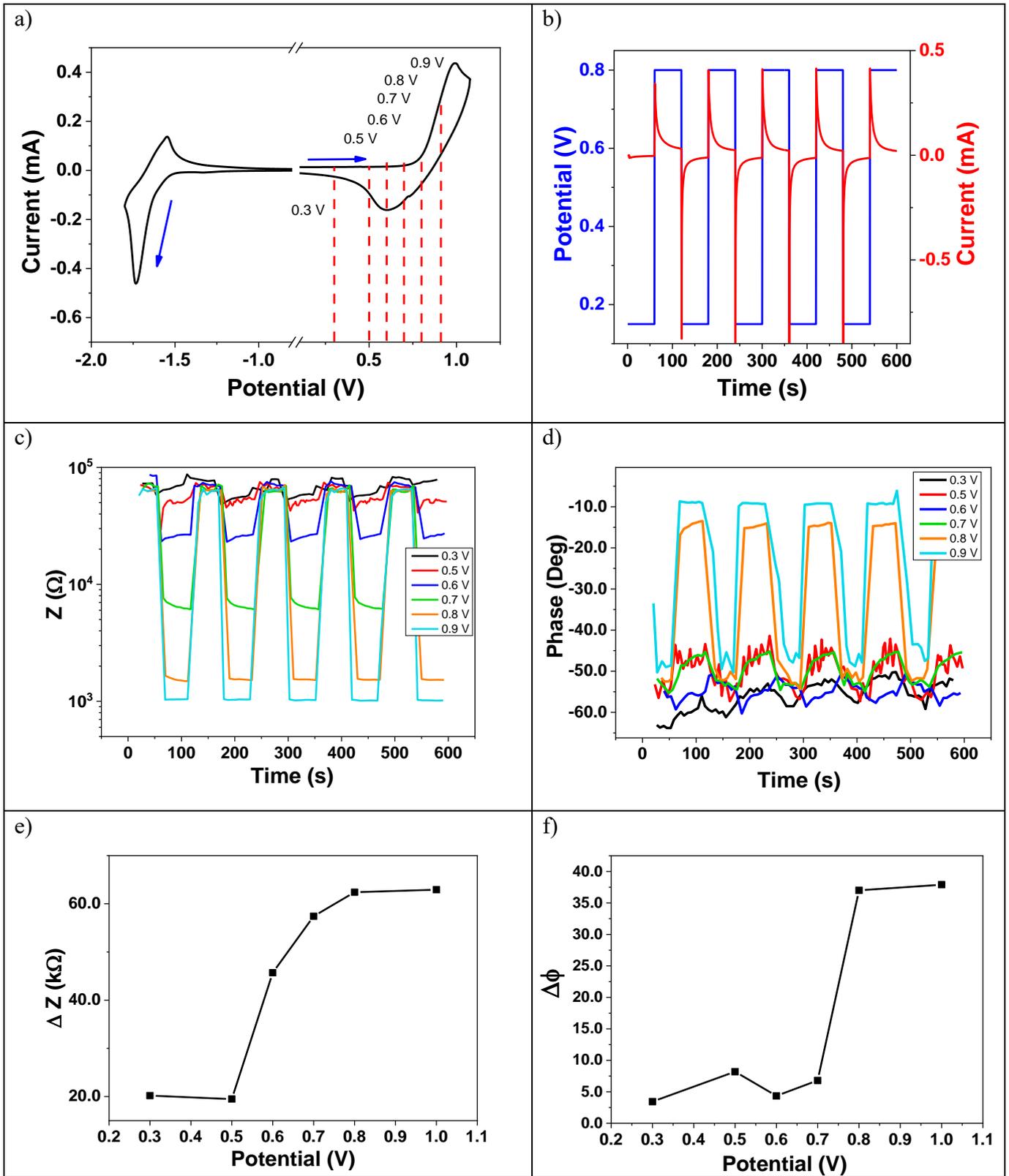

Fig. 2 a) CV curves of a PCPDT-BT film drop-casted on a gold electrode, in a conventional three-electrodes electrochemical cell: GC/PCPDT-BT working electrode, Pt wire the counter electrode, Ag/AgCl/KCl$_{sat}$ the reference electrode, 0.1 M TBATFB in ACN is the base electrolyte, 50 mV s$^{-1}$ potential scan rate. b) Potential vs. t pattern (blue line), applied to the PCPDT-BT in the ECT configuration: compare Figure 1a for details on the experimental set up. Here it is reported as an example the +0.1 to +0.8 V chronoamp (potentiostatic control) time pattern, with a 60 s period. The red line shows the relevant current value. c) and d) Time dependence of in-situ electrochemical impedance spectroscopy measurements, impedance module and phase vs time are plotted, respectively: perturbation signal is a sine wave with 97.7 Hz constant frequency and a 5 mV peak-to-peak potential difference.

*De-doping dynamics*

The WE potential is allowed to relax without any constraint to its final value, under open circuit regime (the electrochemical system is not under any potentiostatic control). The open circuit potential (OCP) is measured in a completely passive mode (OCP means zero current). Here the de-doping dynamics was probed by the simultaneous recording of time series of the OCP in direct current, and by recording impedance vs. time transients. Figure 3 allows for a single glance comparison of the time simultaneous decay of the open circuit potential (where the OCP is the WE to RE potential difference pertinent to the electrochemical system) vs t, impedance module (|Z|) and phase (°) vs. time. In this experiment the initial potential, Eini, of the WE is kept constant for 60 seconds via the potentiostat, then the potential control is lifted and the WE potential is left "free" to relax under zero-current constraint, i.e. OCP conditions. Four different experiments have been recorded by selecting different Eini values: 0.8 (gray dots), 0.9 (red dots), 1.0 (blue dots) and 1.1 (black dots) V, please compare Figure 3. A first striking difference is evident when comparing the d.c. vs a.c. results. In that, the OCP time transient shows clearly three OCP decay domains: the first one characterized by a hyperbolic (1/t) pattern, is followed by an abrupt almost vertical (and substantial) decrease of the OCP as indicated by arrows labelled I in Figure 3a. Then, the steep drop in the OCP is eventually followed by a slowly varying final hyperbolic decay leading to the final equilibrium OCP, marked by arrows labelled II in Figure 3a. Impedance time transients show a qualitatively different pattern, both Z and $\phi$ relax monotonically to a final constant value, please compare Figure 3 b and c. Remarkably, the time at which Z and $\phi$ reach their plateau constant value corresponds to the time of the down-vertical drop in the OCP vs t transients, i.e. arrows labelled I in Figure 3a. A clear indication that the conduction in the PCPDT-BT is related to the WE/solution interface potential in the first part of the OCP transient, then the d.c. and a.c. time patterns reach a time-independent (and potential independent) plateau. The initial decay of the OCP time transients can be effectively simulated by a hyperbolic function, this suggests that the process, in the initial stage, follows a Cottrell diffusion behaviour. Z values reach a constant plateau in a $10^{3.4}$ to $10^{3.7}$ $\Omega$ range, a value substantially smaller than the high-resistivity state (pristine un-doped polymer) of about $10^5$ $\Omega$ (compare in Figure 2c the low-conductivity transients at the +0.1 V potential). The same line of reasoning holds for $\phi$. This is a clear indication that a significative fraction (about the 50%) of the charge injected via electrochemical polarization (doped state) is retained in the polymer even after the instrumental d.c. potentiostatic control is removed. Moreover, the impedance results as a function of time suggest that the PCPDT-BT post-doping state is neatly different with respect to the pristine one, this because the phase of the impedance reach a $-45°$ plateau value, but Z is significative lower than the initial pristine PCPDT-BT value. The latter corresponds to zero conductivity, i.e. the 100 k$\Omega$ in Figure 1c. Thus, the post-doping PCPDT-BT electric state can be described in terms of a R C in parallel discrete components circuit, as it is shown in Figure 3d. Figure 3e shows a cartoon schematizing the pristine un-doped state and the interfacial charge density distribution underlying the initial OCP value. Then, the final state of the whole system features still free to move net-charges within the PCPDT-BT, where essentially the PCPDT-BT strands are seen as a one-dimensional (1D) conductor, simultaneously there is a strong capacitive conductive contribution due to the "molecular"-capacitors made of the single charged polymer strands which are in tight contact. Indeed, the Nyquist spectrum of a R C in parallel discrete components circuit is a semi-circle, featuring a maximum with respect to the real part of the impedance ($z_r$) for $\phi = -45°$ (please compare Figure S8 in the Supporting Information). In addition, the presence of net-charges within the polymer, also after lifting the electrochemical control, is coherent with the difference in the OCP found experimentally, the cartoon in Figure 3 e and f depicts a qualitative picture consistent with such an experimental outcome.

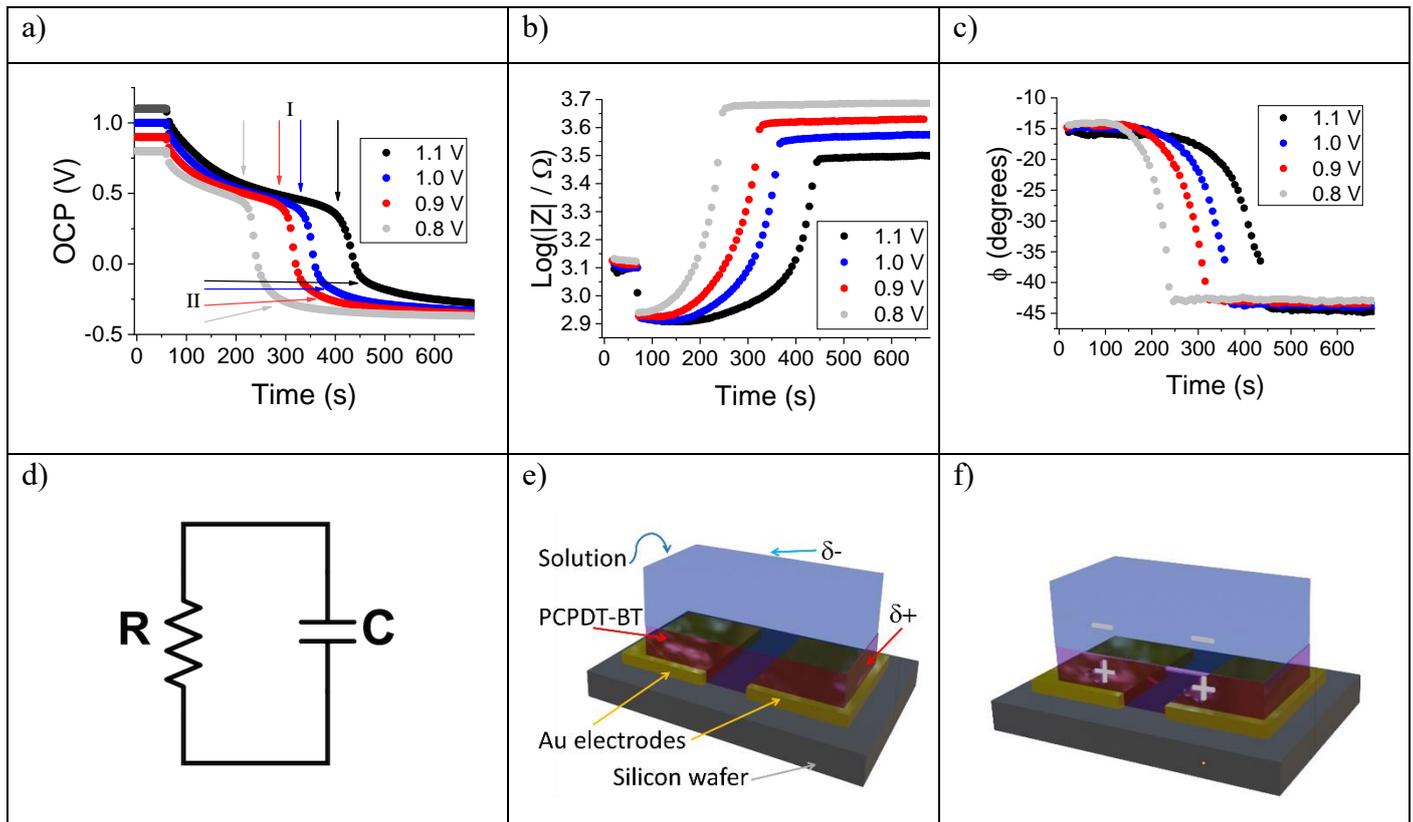

Figure 3. Natural de-doping d.c. and a.c. time transients. a) OCP (where the OCP is the WE to RE potential difference of the electrochemical system) vs time. b) impedance module (|Z|) vs. time. c) and phase (ϕ) vs. time. The initial potential, Eini, of the WE is kept constant for 60 seconds, then the potential control is lifted and the WE potential is left to relax under zero-current constraint, i.e. OCP conditions. Four different experiments have been recorded by selecting different Eini values: 0.8 (gray dots), 0.9 (red dots), 1.0 (blue dots) and 1.1 (black dots) V. d) schematic representation of the polymer equivalent-circuit e) interfacial charge density distribution determining the pristine (un-doped) PCPDT-BT/solution interface OCP value f) charge density distribution determining the PCPDT-BT/solution OCP value after lifting the electrochemical control.

*Polaron IRAV giant response*

Figure 4 shows IR spectra of PCPDT-BT films in both the un-doped and doped states, together with CAMB3LYP/6-311G** theoretical results. Figure 4a sets out, for the sake of comparison, the IR spectrum of a pristine film of PCPDT-BT recorded in the 900 to 4000 cm-1 range, ATR mode using a ZnSe crystal.(26)

Figure 4b shows the theoretical spectrum calculated at the CAMB3LYP/6-311G** level of the theory, a satisfactory agreement between experimental and theoretical results is found, especially concerning the 1000 to 1700 cm-1 range. Figure 4c shows the spectrum of the PCPDT-BT film after exposure to iodine vapour (experimental details can be found in the Supplementary Information), a dramatic change in the IR spectrum is evident, with respect to the pristine one. A broad and swallow peak is centred at about 3000 cm-1 (3333 nm, 0.372 eV), indicated as P1. The latter can be addressed as the "electronic polaron" signature. Nicely corresponding to the broad peak at about 3500 nm in the electronic spectra reported by Koch and Nguyen where the PCPDT-BT chemical doping was obtained by using tris(pentafluorophenyl)borane (BCF) Figure 3d(32) and by Loi and Brabec where the PCPDT-BT chemical doping was obtained by using[6,6]-Phenyl C61 butyric acid methyl ester (PCBM) Figure 2left.(28) A further notable variation between IR spectra in the pristine and doped states is evident at about 1050 cm-1 (9523 nm), where a high intensity peak is present, marked P2 in Figure 4c. The latter intensity, peak P2, is about twice the intensity of the group of peaks centred at around 1400 cm-1, the opposite situation holds in the pristine IR spectrum in Figure 1a. The presence of peak P2 can be assigned to the so called IRAV giant mode, PCPDT-BT vibrational polaron signature. Our results obtained via chemical doping compare well with IR spectra of PCPDT-BT chemically doped reported in the literature.(26–28) Further on, Figure 4d sets out ΔA/A IR spectra of the PCPDT-BT film recorded in-situ: IR spectra recorded applying a constant 0.9 V electrochemical potential (the ΔA/A ratio was calculated as (Abs$_{0.9V}$ - Abs$_{Pristine}$)/Abs$_{Pristine}$). Remarkably the electrochemical doping yields an IR spectrum which is quite similar to that collected via chemical doping. Both the P1 and P2 peaks are present, and even a little more prominent to what is found with chemical doping. Straight comparison with IR spectra obtained by chemical doping allows to assign both P1 and P2 peaks to the electronic and vibrational polaron signature. Eventually note that the sharp oscillations in between the 1200 and 1500 cm-1 range are also consistent with Fano resonances, compare Figure 2left in reference(28). Here, results obtained via chemical doping serve as reference for a tight comparison with the spectra measured by applying an electrochemical bias "in-situ". Thus, there is an evident correlation between cyclic voltammetry results, impedance measurements and IR spectra as a function of the applied electrochemical potential. For potential values larger than 0.7 V, which can be defined as a doping threshold potential, impedance results show a dramatic increase in conductivity, and correspondingly IR spectra showcase the appearance of the "giant IRAV" signature at around 1000 cm-1.(26, 27, 32) Thus, based by comparison of electrochemical results, impedance measurements and IR vibrational measurements, conductivity in the doped state can be assigned to the PCPDT-BT polaron.

| a) PCPDT-BT neutral | b) PCPDT-BT theo |
| --- | --- |

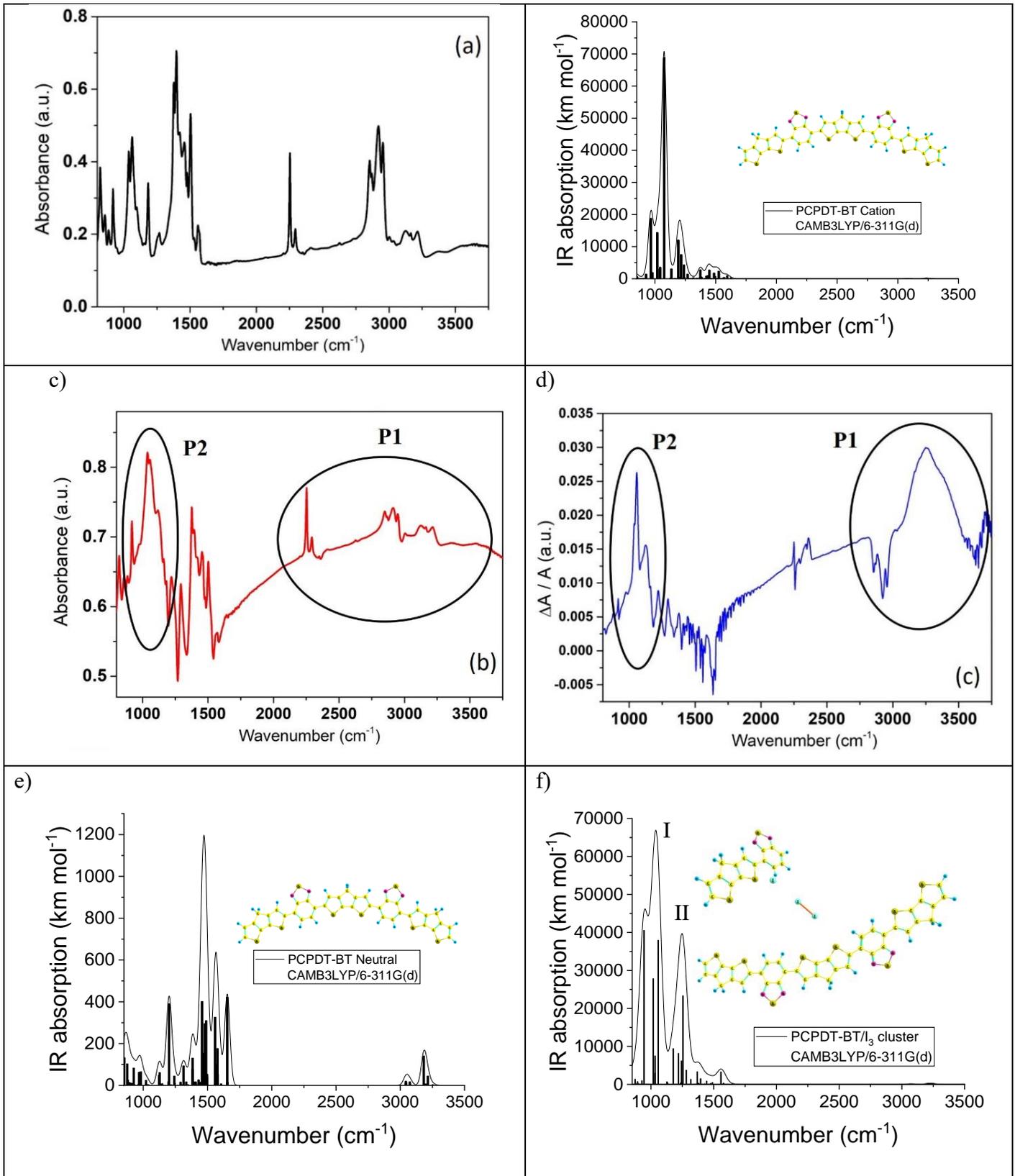

**Figure 4.** Experimental (ATR mode) and theoretical IR spectra of the PCPDT-BT in different pristine and doped states. a) Pristine drop-casted film b) Neutral theoretical spectrum c) Chemically doped by iodine d) Electrochemically doped: 0.9 V. d) Cation theoretical spectrum. e) Theoretical spectrum: PCPDT-BT(2 strands)/I3 cluster. Figure S10 shows vector displacement relevant to the polaron vibrational mode.

Indeed, experimentally it is observed that a dramatic increase (addressed as: "giant IRAV") in intensity of well-defined vibrational modes allows for a very sensitive probing of charged states, with respect to the so-called "electronic polaron-signature".(26) Figure 4 e and f show the computed vibrational spectra for a polymer strand of PCPDT-BT, the insets in Figure 4 e and f represent the PCPDT-BT oligomer and PCPDT-BT(2 strands)/$I_3$ cluster molecular structures. Concerning theoretical results, let us focus on IR spectra in the region of the conjugated backbone vibrations, i.e. the 800 – 1600 cm$^{-1}$ wavelength range. A rather good agreement it is found in the prediction of the two main IR peaks, found at 1000 and 1250 cm-1, indicated as I and II in Figure 4f. Remarkably, theoretical spectra are here reported without any frequency scaling factor. I tight agreement with experimental results, a clear-cut variation is evident between the theoretical IR spectra of the neutral copolymer oligomer and that of both the cation and the PCPDT-BT(2 strands)/$I_3$ cluster as reported in Figures 4 e and f. Particular care must be paid when comparing the ordinate full scale axis: the vibrational polaronic giant IR mode is 58 times larger than the largest intensity peak of the neutral PCPDT-BT neutral oligomer  (obtained dividing the IR intensity full scales of the charge transfer cluster with respect to the neutral oligomer strand, that is 70000/1200 in Figures 4 f and b, respectively). Indeed, the giant IRAV mode in the PCPDT-BT cation oligomer (the process underlying P2 in both the chemical and electrochemical doping) is found at 1050 cm-1, Figure 4g shows the relevant displacement vectors, and remarkably these facts and figures cope quite well with the giant mode calculated for the PCPDT-BT(2 strands)/$I_3$ cluster where the highest intensity IR peak is found at 1045 cm-1 featuring a quite similar pattern of IR mode displacement vectors, as can be verified by comparing Figure 4 g and h. Essentially, in the calculated results, the giant mode is found ranging between 1000 and 1080 cm-1. The giant vibration can be classified as a waggling mode involving both a bending of the thiophenes C-C-H bond and stretching of the C-C thiophene bond opposite to the sulphur atom.

**Conclusions**

This work demonstrates the implementation of a device were a solid-state transistor (source/drain/gate) electrical layout is combined with an electrochemical system. To this purpose a well-studied organic semiconductor, PCPDT-BT, is exploited as the "trait d'union" acting as both the gate in the solid-state device and working electrode in the electrochemical cell. The electronic state of the gate (doped/un-doped) is controlled via electrochemical polarization. The source to drain conductivity is measured via impedance spectroscopy. The conductivity in the doped state can be unambiguously assigned to the polaron induced in the PCPDT-BT upon electrochemical polarization, as it is shown by recording ATR IR spectra in both electrochemical "in-situ"/"in-operando" conditions and via chemical doping (iodine doping), as well as by comparison of experimental and theoretical results from the literature. The original electrochemical transistor concept here proposed can find applications in different fields of both science and technology, and coupled to different detecting techniques like EPR or Hall effect measurements.(33, 34) In particular it appears a promising sensor device, because its sensitivity to analytes in solution can be probed by direct control obtained by the electrochemical potential of the undoped/doped electronic states of the organic semiconductor. In addition, beyond the conductivity (i.e. in the end an amperometric based measurement) the phase of the impedance shows a very high sensitivity to the organic semiconductor doping state, and could be exploited to implement a capacitor based sensor device. Remarkably, due to the polymer long structure of PCPDT-BT, our integrated impedance, vibrational and theoretical outcome provides a self-consistent picture where the PCPDT-BT conductivity is well described in terms of a 1D polaron model.(35–37)

**Methods**

*Chemicals*

Merck Dichloromethane (DCM) ≥ 99.5%, Merck Acetonitrile solvent (ACN) ≥ 99.5%, Merck Tetrabutylammonium tetrafluoroborate $(CH_3CH_2CH_2CH_2)_4N(BF_4)$ (TBATFB) 99%, Sigma Aldrich PCPDT-BT and Sigma Aldrich Iodine ($I_2$) ACS reagent ≥ 99.8% were used without further purification.

*Electrochemical transistor*

Electrochemical gating measurements on the electrochemical transistor system were conducted using a two potentiostats setup, as depicted in Figure 1a. An Autolab PGSTAT 128N potentiostat was employed to apply the gate potential (WE) in direct current at the WE, while the alternated current impedance was measured using a CHI660A potentiostat between the source and drain pads). In a conventional electrochemical cell configuration, a glass substrate was utilized, and three 100 nm-thick Au pads (with a 10 nm adhesion layer of Chromium) were evaporated onto it. The PCPDT-BT polymer was drop-cast between these pads, serving as the gate. Two of the three pads coated with PCPDT-BT were then connected as the working electrode (WE) and counter electrode (CE) to the CHI660A potentiostat for impedance measurements, functioning as the source and drain, respectively. The third pad acted as the WE, and was connected to the potentiostat driving the electrochemical part of our complex set up, which applied the DC gating voltage. This allowed impedance spectra to be recorded as a function of the applied gating voltage.(38, 39) The thickness of the PCPDT-BT film is the range of 200 to 500 nm and can be controlled by adjusting the concentration of the PCPDT-BT solution and the choice of solvent. Modest variations in the thickness of the drop-casted film do not significantly affect conductivity, the latter is primarily due to the 2D PCPDT-BT film in direct contact with the source and drain electrodes.

*In-situ in-operando FTIR spectroelectrochemistry*

The doped state of a drop-casted film of PCPDT-BT (approximately 1 μm thick) is electrochemically controlled by applying a potential bias via a glassy carbon (GC) or Au termials in mechanical and electrical contact with the PCPDT-BT. The latter is connected to the potentiostat as the working electrode (WE). The electrochemical cell containing the electrolyte (in this case, ACN with 0.1 M TBATFB as the supporting electrolyte) is physically obtained by exploiting the side-walls of the accessory for supporting the ATR crystal (Gateway HATR System, SPECAC, 8 internal reflections, ZnSe crystal). This arrangement creates a small cavity capable of holding around 4 ml of electrolyte solution, sufficient for establishing contact with a Pt wire counter electrode (CE), and a Ag/AgCl reference electrode (RE), compare the scheme show in Figure 1b. A figure showing explicitly the experimental setup for FTIR spectro-electrochemistry measurements, conducted to investigate the polaronic state of PCPDT-BT, is presented in the Supporting Information, Figure S9.

*Theoretical calculations.*

Ab initio molecular orbital calculations were performed using the ORCA and NwChem suite of programs.(40, 41) Please note that both the PCPDT-BT single molecule and PCPDT-BT$_n$/I$_m$ clusters are characterized by an extremely large number of conformational degrees of freedom. Both structural (PCPDT-BT strand length, as well as neglecting the presence of later saturated carbon side-chains) and conformational analysis took advantage of previously quite accurate studies on PCPDT-BT vibrational spectra from this lab.(26, 27) Eventually, CAM-B3LYP/6-311G** level of the theory was selected as a compromise between accuracy and calculation costs, giving a quite accurate agreement with the experimental vibrational spectra. Indeed, the use of 3-21G basis set was able to yield qualitatively correct results, giving due count of both the electronic an5d vibrational polaron signature.

**Acknowledgements**

Financial support is gratefully acknowledged from the Ministry of University and Research (MUR), PRIN 2022 cod. 2022NW4P2T "From metal nanoparticles to molecular complexes in electrocatalysis for green hydrogen evolution and simultaneous fine chemicals production (FUTURO)", PI Dr. Francesco Vizza, from Fondazione di Modena, Fondo di Ateneo per la Ricerca Anno 2023, linea FOMO, Progetto AMNESIA and from Consorzio Interuniversitario Nazionale per la Scienza e Tecnologia dei Materiali (INSTM), fondi triennali: "INSTM21MOFONTANESI".

# Supporting Information

The Electrochemical Transistor: a device based on the Electrochemical control of a polymer Polaronic state. The PCPDT-BT as a case study.

Andrea Stefani[1], Agnese Giacomino, Sara Morandi, Andrea Marchetti, Claudio Fontanesi.*

[1]*Department of Physics, (FIM), Univ. of Modena, Via Campi 213/A, 41125 Modena, Italy.*

*Department of Chemical and Geological Science, DSCG, Univ. of Modena, Via Campi 103, 41125 Modena, Italy.*

*Department of Engineering "Enzo Ferrari", (DIEF), Univ. of Modena, Via Vivarelli 10, 41125 Modena, Italy.*

*National Interuniversity Consortium of Materials Science and Technology (INSTM), Via G. Giusti 9, 50121 Firenze (FI), Italy.*

**Index:**

1) Electrochemical transistor architecture

2) Impedance cross-check

- Blank, undoped polymer conduction cross-check.
- Impedance time transients as a function of frequency.
- Chemical doping.

3) Impedance data elaboration.

4) Details of the "in-situ"/"in-operando" spectroelecrochemical IR experimental set up

5) Polaron vibrational mode displacement vectors.

*1) Electrochemical transistor architecture*

Figure S1a illustrates the architecture of the substrate with 3 conductive pads for the electrochemical transistor assembly. The channel between the source and drain has a width of 0.5 mm. For the sake of comparison of non-conductive substrate material, also a glass microscope cover slip was used in screening experiments, anyhow devices with glass or silicon wafer substrate yielded the same results.

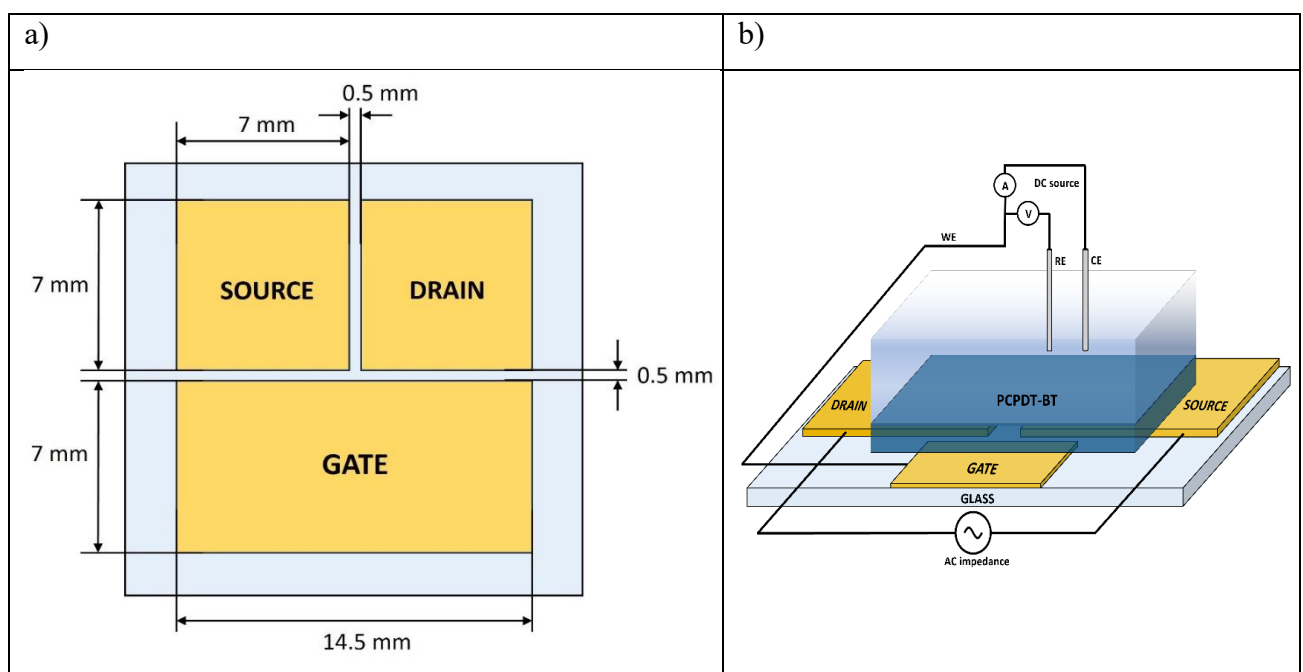

Figure S1. a) Conductive surface layout of the electrochemical transistor device. b) Same device as in Figure 1a main manuscript, but the isolating substrate is a microscope cover slip.

*2) Impedance cross-check*

*Blank, undoped polymer conduction cross-check.*

Figure S2 shows cross check impedances, in the 1 Hz to 100 kHz frequency range, recorded between the device source and drain contacts (see Figure 1a of the main manuscript for the details), of the PCPBT-DT films just dropcasted on the electrochemical transistor device, Figure 1a main manuscript. Figure S1a shows the Bode plot of the impedance in air. Figure S2b shows the Bode plot of the impedance of the device with the PCPDT-BT dropcasted and the device is assembled in the electrochemical cell, that is the device/PCPDT-BT/solution interface, just prior of applying the contacts for the electrochemical polarization. To obtain a reasonably clean impedance spectrum the pea-to-peak voltage amplitude was increased to 0.1 V. Of course, impedance spectra approaching the d.c. limit (i.e. the 0 Hz frequency limit) get more and more scattered, below 1000 Hz the Bode plot gets essentially a random collection of points. In Figure S2 a and b the phase is virtually constant, and close to -90°, for frequencies larger than 1000 Hz, this indicate an almost purely capacitive behaviour, implying that Z tends to infinite as the frequency tends to zero: for values lower than 1000 Hz, although not quantitatively precise, Z values are much larger than 0.5 GΩ.

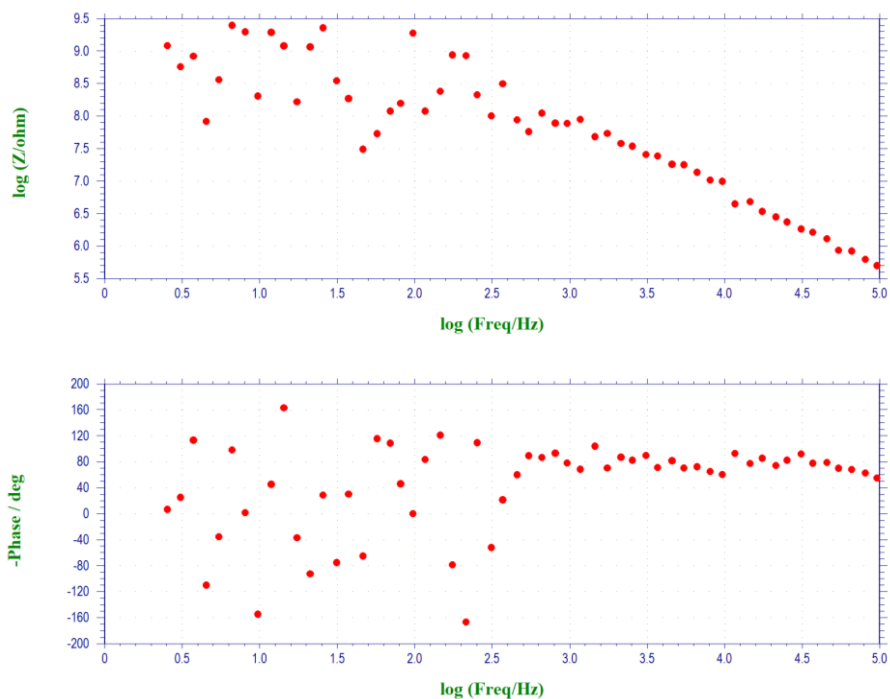

Figure S2a. Pristine (un-doped) PCPDT-BT dropcasted on the device.

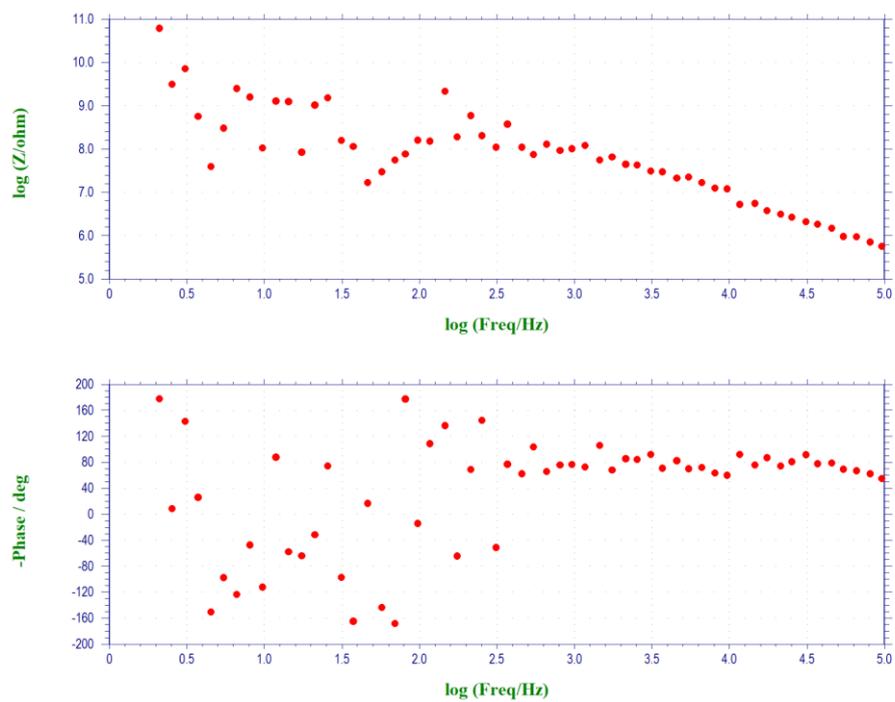

Figure S2b. Pristine (un-doped) PCPDT-BT dropcasted on the device in contact with the electrolyte solution, ACN 0.1 M TBATFB.

*Impedance time transients as a function of frequency.*

Impedance vs time transients were recorded as a function of the perturbating sinusoidal signal, with the same potential vs time chronoamperometry potential steps. Impedance module vs. time and phase vs. time transients as a function of the perturbation potential frequency. The PCPDT-BT state is driven under potentiostatic regime, i.e. chrono amperometry, switching the potential between 0.0 V, kept constant for 10 s, and 0.8 V, constant for 10 s.

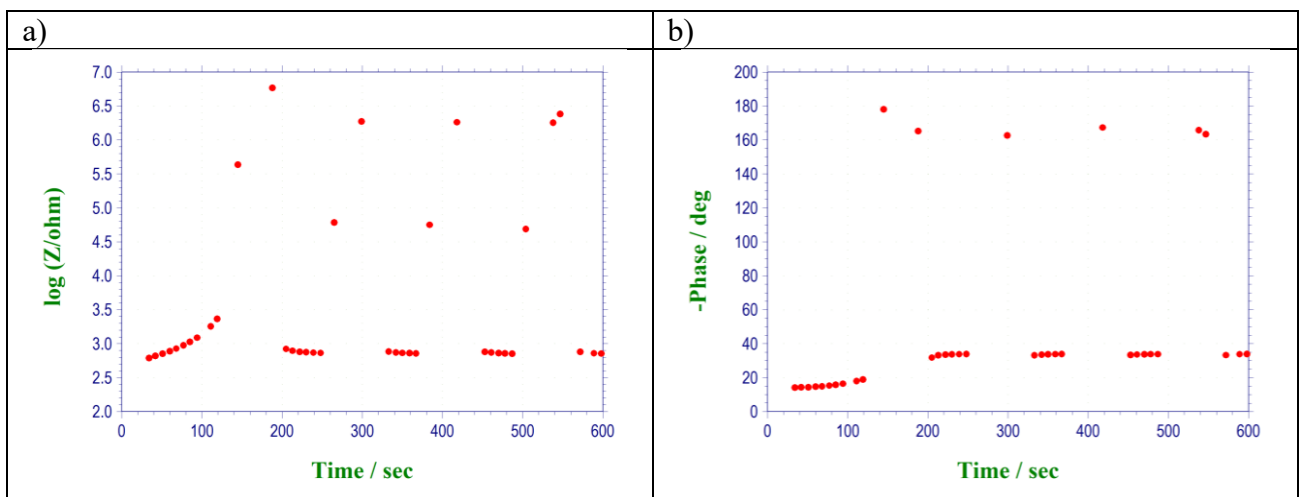

Figure S3. Electrochemical transistor, impedance as a function of time, 1.772 Hz, 5 mV peak-to-peak amplitude. a) Absolute value as a function of time. b) Phase as a function of time.

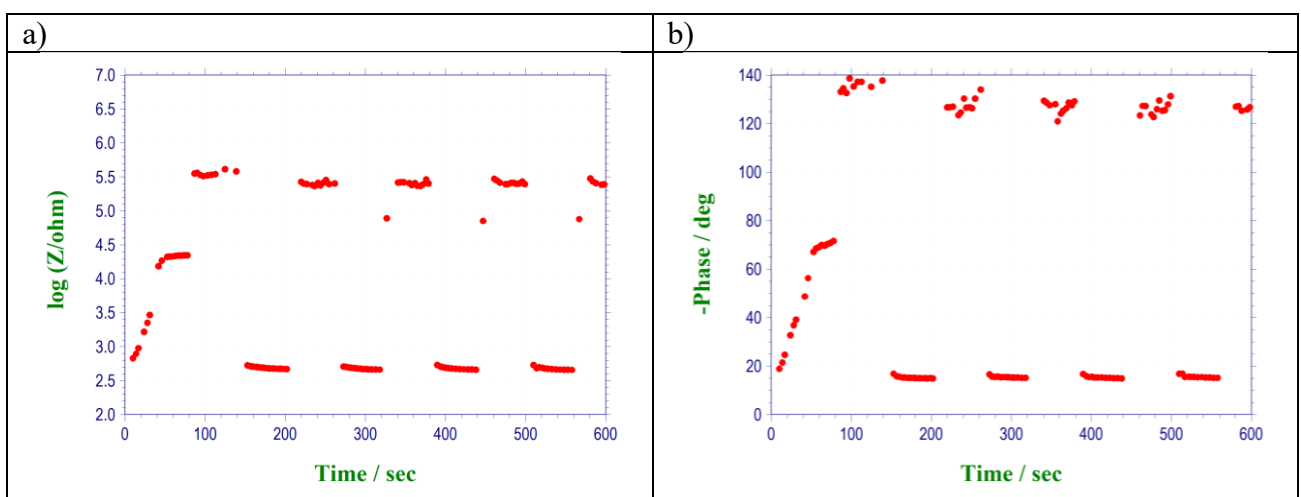

Figure S4. Electrochemical transistor, impedance as a function of time, 17.2 Hz, 5 mV peak-to-peak amplitude. a) Absolute value as a function of time. b) Phase as a function of time.

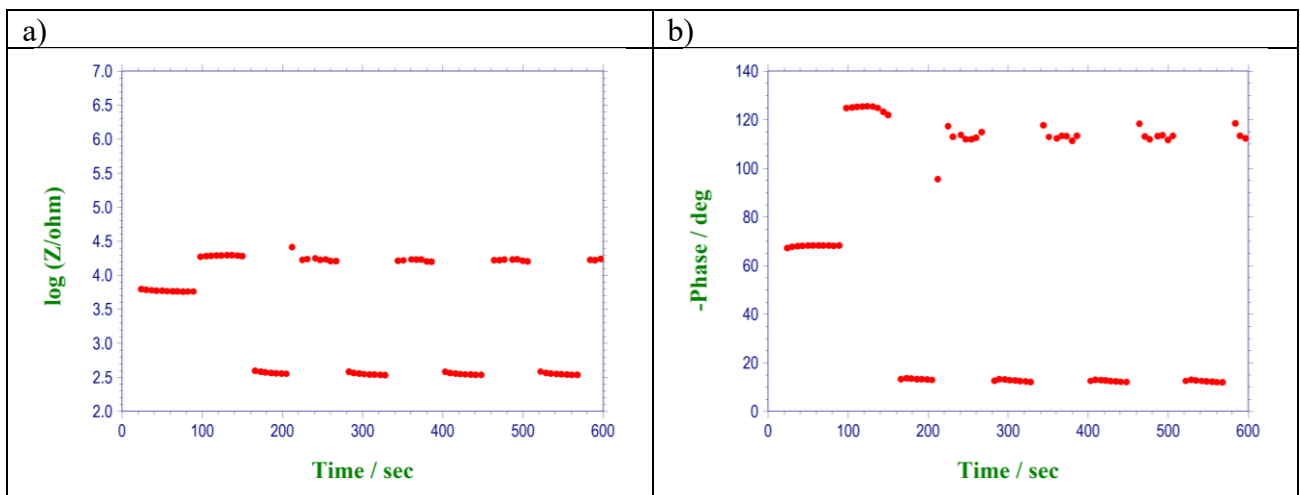

Figure S5. Electrochemical transistor, impedance as a function of time, 123 Hz, 5 mV peak-to-peak amplitude. a) Absolute value as a function of time. b) Phase as a function of time.

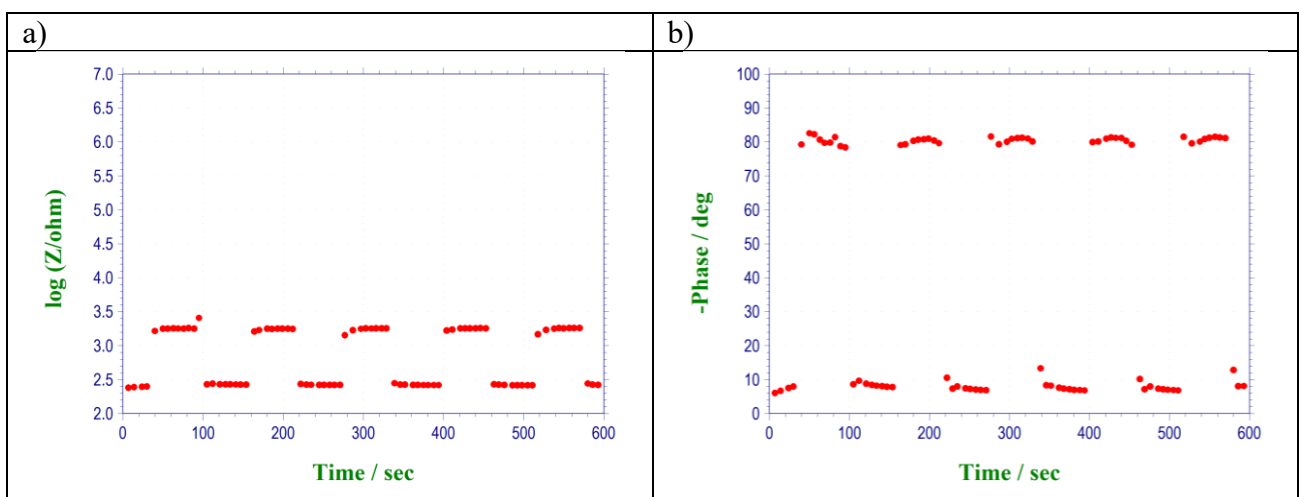

Figure S6. Electrochemical transistor, impedance as a function of time, 962 Hz, 5 mV peak-to-peak amplitude. a) Absolute value as a function of time. b) Phase as a function of time.

*Chemical doping.*

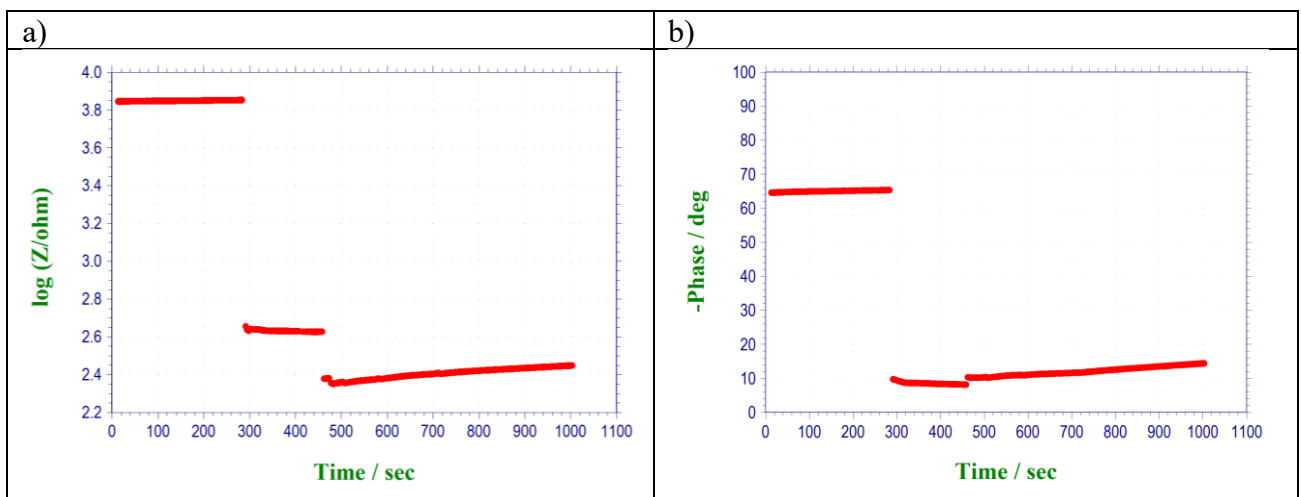

Figure S7. PCPDT-BT chemically doped with iodine, impedance as a function of time, 18.1 Hz, 5 mV peak-to-peak amplitude. a) Absolute value as a function of time. b) Phase as a function of time.

3) Impedance data elaboration.

As assumed in the main manuscript, the electrical behaviour of the PCPDT-BT, following the remove of the electrochemical potentiostatic control, can be represented as a R C in parallel electric circuit. Then, the relevant analytical impedance function is:

$$Z(\omega) = \frac{R}{1+(\omega CR)^2} - j\frac{\omega CR^2}{1+(\omega CR)^2}$$

$$Z(\omega) = \frac{R}{1+(\omega CR)^2} - j\frac{\omega CR^2}{1+(\omega CR)^2} \qquad \text{Eq.1}$$

Considering the Nyquist representation, then:

$$z' = \frac{R}{1+(\omega CR)^2} \qquad \text{Eq.2}$$

and

$$z'' = -\frac{\omega CR^2}{1+(\omega CR)^2} \qquad \text{Eq.3}$$

The relevant Nyquist plot, the implicit plot of $z'(\omega)$ vs. $z''(\omega)$, is:

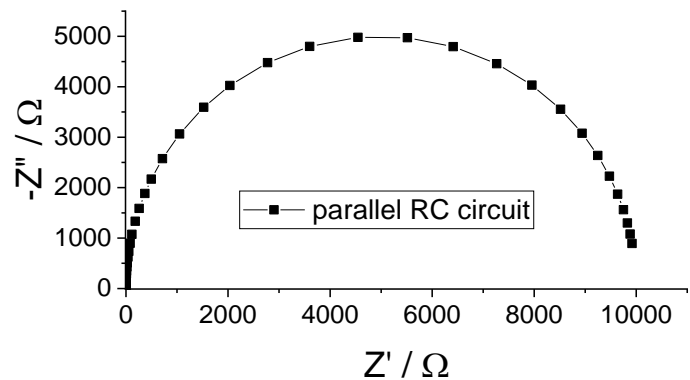

Figure S8. Implicit, Nyquist, plot representing $z'(\omega)$ vs. $z''(\omega)$ for an ideal R C parallel circuit.

Thus, when $\phi = -45°$, $z' = z''$ (compare Figure S8). Moreover, since $\omega = 2 \times \pi \times f$ (with $f = 97.7\ Hz$) and $|Z| \cong 10^{3.5}$, i.e. the absolute value of Z at the plateau (please compare Figure 3b I the main manuscript), a value of $C \cong 1\mu F$ can be estimated.

*4) Details of the "in-situ"/"in-operando" spectroelecrochemical IR experimental set up*

As it is shown in Figure 1b main manuscript, a crucial role is played by the ATR Specac Multi-reflection HATR accessory: a eight reflections ZnSe crystal is embedded in a frame able to act as a shallow, 3 mm depth, trough thus allowing the electrolyte solution to be in contact with the WE (the PCPDT-BT film, the electrochemical contact, to the potentiostat, is implemented via a glassy carbon electrochemically inert rod). Platinum and silver wires, CE ad RE respectively, allow for the electrochemical control. The SPECAC HATR accessory was assembled within a Vertex 70 FTIR spectrophotometer, Figure S8 low right, the electrochemical control is obtained via the Autolab PGSTAT 128 N, Figure S8 top middle.

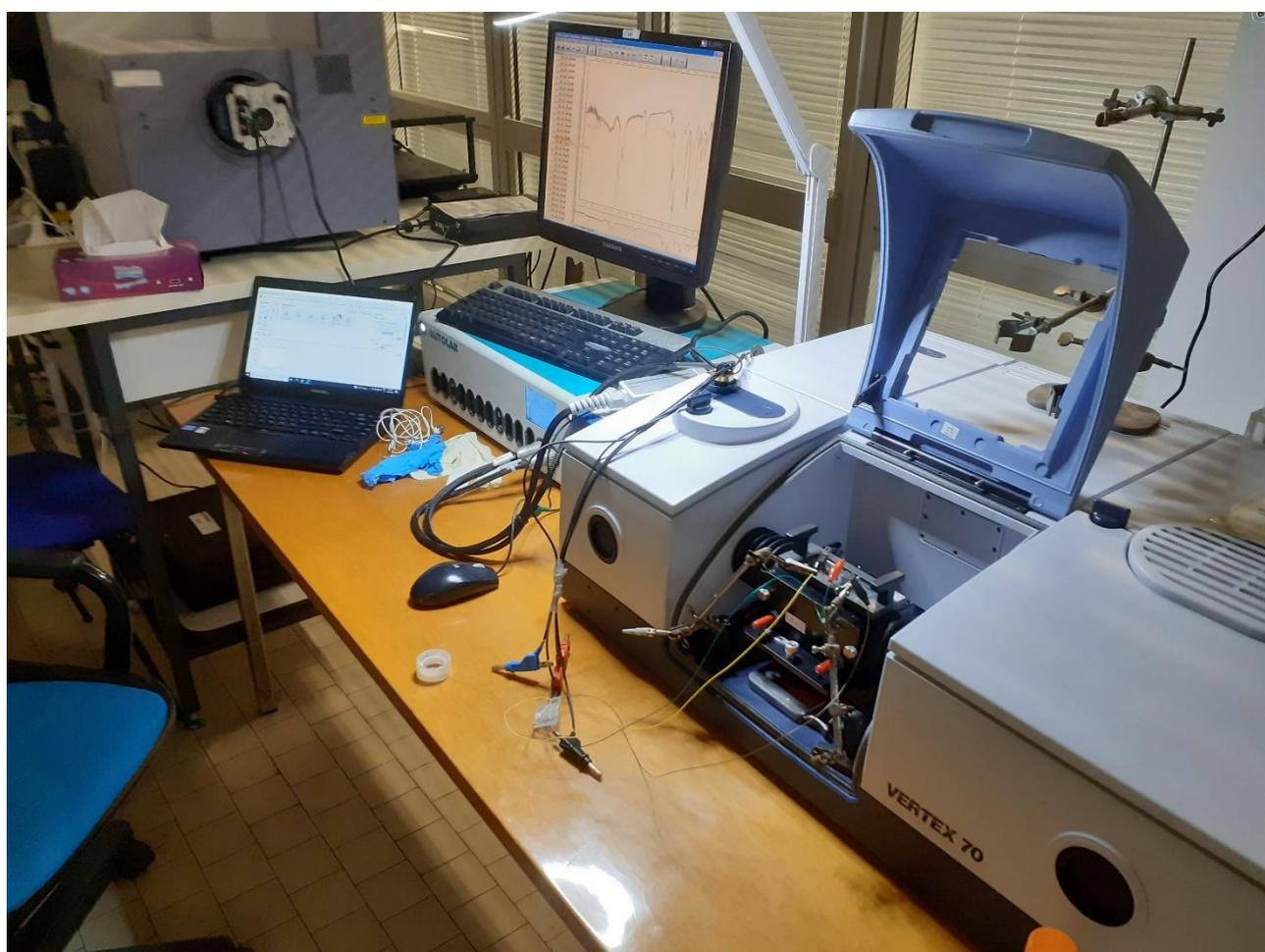

Figure S9. Picture of the experimental arrangement implemented for recording "in-situ"/"in-operando" spectro-electrochemical IR spectra, under real-time potentiostatic control.

5) Polaron vibrational mode displacement vectors.

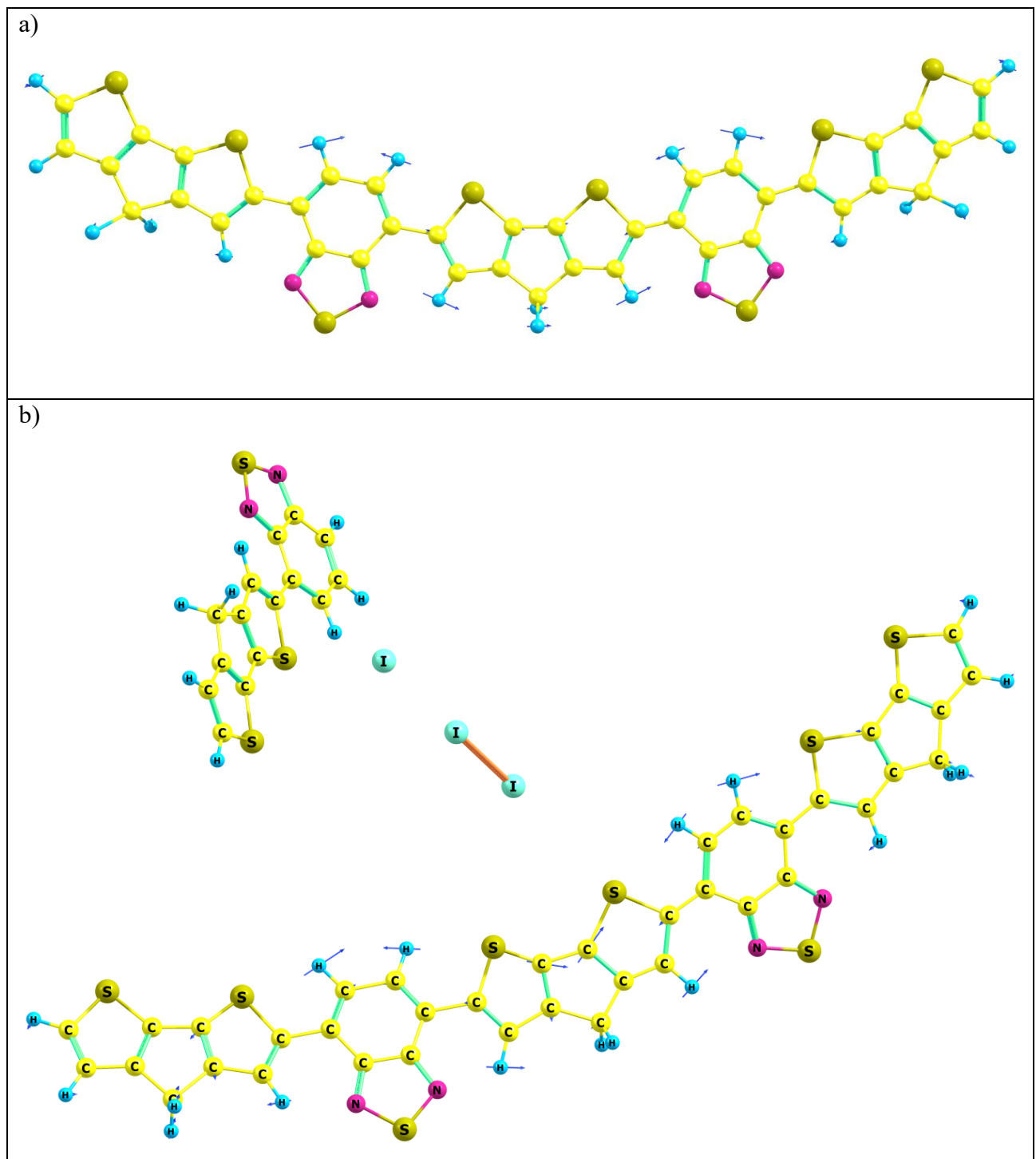

Figure S10. Theoretical IR spectra. a) Cation: displacement vectors, mode number 124, wavenumber 1076 cm$^{-1}$ b) PCPDT-BT(2 strands)/I3 cluster: displacement vectors, mode number 185, wavenumber 1057 cm$^{-1}$